\documentclass[doc]{apa6}

\usepackage[utf8]{inputenc}
\usepackage[]{geometry}
\usepackage{nth}
\usepackage{natbib}
\usepackage{amsmath,amssymb}
\usepackage{graphicx}
\usepackage{subcaption}
\usepackage{wrapfig}
\usepackage{xcolor}
\usepackage{soul}
\usepackage{doi}
\usepackage{multirow}
\usepackage{listings}

\title{Locally Learned Synaptic Dropout for Complete Bayesian Inference}
\shorttitle{Locally Learned Synaptic Dropout}
\fourauthors{Kevin McKee}{Ian Crandell}{Rishidev Chaudhuri}{Randall O'Reilly}
\fouraffiliations{University of California, Davis}{Virginia Tech}{University of California, Davis}{University of California, Davis}

\abstract{
The Bayesian brain hypothesis postulates that the brain accurately operates on statistical distributions according to Bayes' theorem. 
The random failure of presynaptic vesicles to release neurotransmitters may allow the brain to sample from posterior distributions of network parameters, interpreted as epistemic uncertainty.
It has not been shown previously how random failures might allow networks to sample from observed distributions, also known as aleatoric or residual uncertainty.
Sampling from both distributions enables probabilistic inference, efficient search, and creative or generative problem solving. 
We demonstrate that under a population-code based interpretation of neural activity, both types of distribution can be represented and sampled with synaptic failure alone. 
We first define a biologically constrained neural network and sampling scheme based on synaptic failure and lateral inhibition.
Within this framework, we derive dropout based epistemic uncertainty, then prove an analytic mapping from synaptic efficacy to release probability that allows networks to sample from arbitrary, learned distributions represented by a receiving layer.
Second, our result leads to a local learning rule by which synapses adapt their release probabilities.
Our result demonstrates complete Bayesian inference, related to the variational learning method of dropout, in a biologically constrained network using only locally-learned synaptic failure rates.
}

\begin{document}
\maketitle
\clearpage

\section{Introduction}
The \emph{Bayesian Brain} hypothesis has led to a number of important insights about neural coding in the brain \citep{knill2004bayesian, friston2010free, friston2012history, pouget2013probabilistic, lee2003hierarchical} by characterizing neural representation and processing in terms of formal probabilistic inference and sampling.  
Furthermore, the introduction of related probabilistic representations and sampling processes in modern deep learning \emph{variational} models has led to improved performance on a range of different tasks \citep{zhang2018advances, blei2017variational, kingma2013auto, detorakis2019inherent}.
The widely-used \emph{dropout} technique in deep learning can be seen as a form of variational inference and sampling \citep{srivastava2014dropout, gal2016dropout} with direct analogy to the random failure of synapses in the brain.
This link has led to biologically-motivated models of variational deep learning that use network weight dropout to simulate synaptic failure \citep{mostafa2018learning, wan2013regularization, neftci2016stochastic}.

In this paper, we build on these and other recent findings in machine learning and neurobiology to show how the brain can accurately represent the two primary components of probabilistic inference, distributions of observed data and distributions of unobserved values (such as model parameters), with the single, biologically established mechanism of synaptic failure.
In the first section, we introduce relevant concepts of probability and review recent developments.  
Next, we propose a theory of neural sampling by synaptic failure and lateral inhibition and show that under that theory, network weights can be analytically mapped to transmission probabilities that result in unbiased samples from arbitrary distributions.
Our result further leads to a local, biologically-realizable learning rule for synaptic failure probabilities, consistent with recent evidence that the rate of synaptic failures appears to be under adaptive control.
Finally, we use simulations to demonstrate complete posterior sampling in an abstracted network using only locally-learned transmission probabilities.

\subsection{Probabilistic reasoning}
To introduce the relevant concepts, consider a simple network that learns about the sizes and weights of different dog breeds.
Any particular sample dog has a well-defined size, but across the population, there is a distribution of sizes.
A \emph{probability distribution} assigns likelihood to each size.
If we consider only a particular breed, then we would obtain a likelihood from the \emph{conditional distribution} of size given that breed. 
It would be useful for a neural system to be able to represent the conditional distributions of sizes, rather than just a central tendency such as the most frequently observed size.
Conditional distributions, as a one-to-many mapping, allow the neural system to explore possibilities and plan accordingly, for instance, when designing dog houses and toys for the most likely range of end users.


There are two primary sources of variance or uncertainty in the target distribution (e.g., dog sizes): the distribution of the data (e.g., the observed sizes across all dogs of a given breed), and the distribution of the parameters in a model of that data (e.g., the network weights or synapses linking size to breed).
Accordingly, any prediction about the size of a particular dog must come in the form of a distribution that draws from both.
The first type of variance is often called the \textit{residual} variance in classical statistics because it refers to observed variation in the dependent variables that is left over after conditioning them on the independent variables. It is called the \textit{risk} in machine learning in reference to a model's probability of misclassification despite converging on a solution in training, and more recently as the \textit{aleatory} (i.e, random) uncertainty in variational deep learning.
Conversely, uncertainty in parameters is known in classical statistics, as the \textit{sampling error} or by its inverse, precision, and serves as the basis for drawing scientific inferences from data.
In recent variational deep learning literature, parameter imprecision has been called the \textit{epistemic uncertainty}.\footnote{This use of epistemic uncertainty to refer only to parameter imprecision is a misnomer in that the residual distribution also depends upon epistemic degrees of freedom, namely what data are chosen and how the models are specified.
More broadly, epistemic uncertainty includes problems of measurement, experimental design, interpretation of model results, and many other sources that are unlikely to be represented in the model itself.
\cite{o2016weapons} cites wide-spread, misguided confidence in machine learning models as a source of systemic bias and inequity in its societal impacts, which are not likely to be resolved by variational methods alone.} 
In Bayesian modeling, the goal is typically to estimate the posterior distribution of the parameters given data and a prior distribution, which can then be used to draw inferences about the unknown quantities in question. 

As residual variance reflects observations that the model cannot ``explain away,'' learning to sample from the residual distribution can allow a network to explore the range of events that are known to be possible.
To the extent that different regions of a distribution lead to markedly different processes in subsequent layers, and ultimately different behavioral outcomes, activation of only the mean value or other central tendency will severely limit the capabilities of the network.

Parameter uncertainty, as one form of epistemic uncertainty, produces additional noise that propagates through the network into inferences.
While we might wish to eliminate noise and be maximally certain in any situation, parameter uncertainty itself likely serves the human brain the same way it serves scientific inference: we can avoid drawing premature conclusions and acting too soon. 
Furthermore, whereas the conditional distribution of the data is a kind of record of previous observations, the uncertainty associated with unobserved parameters can extend indefinitely beyond the bounds of previous or future experience.  
For instance, adults learn that there is a finite range of colors and patterns in dogs' fur, and this knowledge only becomes more bounded with additional experience.
Young children, however, may readily imagine dogs with blue, red, or green fur, with less concern for plausibility.
In this way, the process of sampling from what we call epistemic uncertainty may hold implications for the search space of human imagination.
While adults grow to be less interested in outlandishly colored dogs, we remain no less imaginative about other unknown aspects of nature.

While the relationship between a conditional distribution of data and any associated parameter distributions can be derived analytically for simple, parametric models like linear regression, doing so is difficult or intractable for models in general, requiring numeric estimation instead.
In classical inference with maximum likelihood estimation, the Fisher information is computed from the matrix of parameter second derivatives with respect to the likelihood function and provides estimates of the complete parameter covariance matrix \citep{lehmann2006theory}.
In Bayesian modeling, Markov Chain Monte Carlo is used to draw samples from the posterior distribution of the parameters, which can then be used to calculate means, modes, and variances.

\paragraph{Variational inference}
Deep learning models are too complex for exact analytic or numeric inference to be practical or even possible, so recent developments have primarily focused on \emph{variational inference} as an alternative.
Variational inference refers to the use of greatly simplified, approximate distributions to draw inferences from complex models.
Variational neural networks were pioneered with the Boltzmann machine \citep{hinton2007boltzmann}, Helmholtz machine \citep{dayan1995helmholtz}, and their later generalization, the variational autoencoder \citep[VAE,][]{kingma2013auto} and have since been further generalized to networks of many kinds \citep{zhang2018advances}, including networks that imitate neurobiology \citep{mostafa2018learning, neftci2016stochastic}.
In these models, sampling from the variational distributions is used to approximate an intractable integral in the objective function, to learn regularized solutions, and to generate plausible out-of-sample realizations from the learned, latent distributions.
These models generally rely on several mechanisms that are not biologically plausible, including negative weight values, parametric distributions, and global objective functions, typically a KL-divergence function.

\paragraph{Dropout}
Recent innovations in sampling in machine learning have focused on dropout, the random masking of network activations \citep{srivastava2014dropout}, weights \citep{wan2013regularization}, or both, as a form of variational inference \citep{labach2019survey, neftci2016stochastic}.
\cite{gal2016dropout} demonstrated that minimizing the KL-divergence of a Binomial-Gaussian approximation of each network weight to a standard normal prior is functionally equivalent to $\mathcal L^2$ regularization.
As regularization implies a distribution of the parameters within which they may be shifted without incurring excess error, the dropout algorithm allows one method of sampling from that distribution, and hence sampling from distributions of the subsequent node activations that condition on those parameters.
Under that application, the dropout distribution represents uncertainty in the network weights, which then propagates to the activations and final predictions.
To date, it remains to be demonstrated and implemented that weight dropout can be used to sample from the conditional distribution of the data.

\subsection{Biological neural networks}
\begin{figure}[!ht]
    \centering
    \includegraphics[width=\linewidth]{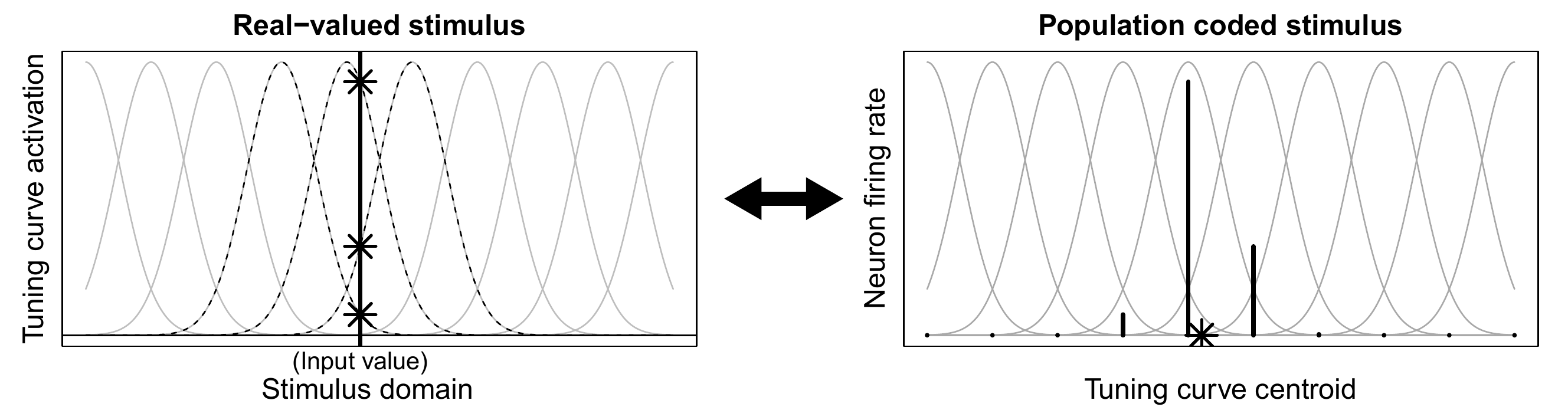}
    \caption{Dual representation of a stimulus as a real-valued number, the position of the vertical line on the left, and as neural firing rates, the heights of the lines on the right. Gray background curves represent the tuning curves of the sensory receptors. Asterisks shows how each representation is understood in the space of the other.}
    \label{fig:popcode1}
\end{figure}

The critical question addressed in this paper concerns how a population of neurons represents distributions and propagates uncertainty toward the end goals of action and perception.
Although it is possible to consider individual neurons as representing entire distributions, it is more likely that the relevant information is distributed across populations of neurons, in a \emph{population code} \citep{zemel1998probabilistic}.
Population codes can be understood as transformations of an input to a set of neural activations according to the neurons' \textit{tuning curves} (Figure \ref{fig:popcode1}).
A widely-used tuning curve is the Gaussian function, which has a central, preferred value of the stimulus, and a width representing the range of stimuli that activate the neuron to a lesser degree.
Input values closer to the center of a Gaussian tuning curve cause the neuron to fire more frequently.
By taking the the point corresponding to the maximum of the weighted sum of many neurons' firing rates, and hence, the heights of their respective tuning curves, values of the inputs can recovered.

\begin{figure}[!ht]
    \centering
    \includegraphics[width=0.7\linewidth]{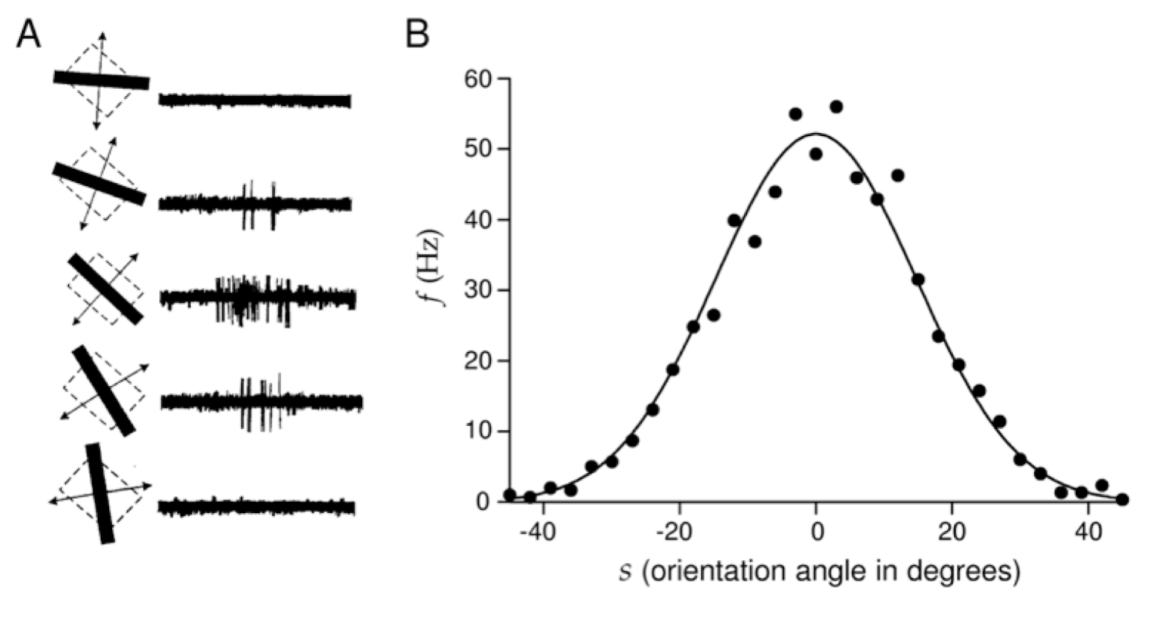}
    \caption{Gaussian kernel population code for visual orientation.  Each neuron has a tuning curve around a preferred orientation, and the entire population of neural activity encodes the current orientation as a weighted average of the neuron activity times its preferred value \citep{hubel1968receptive}.}
    \label{fig:tuningcurve}
\end{figure}

Such codes are well-established in the brain, for example in the population coding of visual orientation angle, or motor variables such as velocity or angle of a saccade \citep{Albright84,GeorgopoulosSchwartzKettner86,SommerWurtz00}. 
As shown in Figure \ref{fig:tuningcurve} \citep[taken from][]{hubel1968receptive}, the orientation of a line is represented by the firing rates of several neurons with distinct, preferred values.
If the receiving layer, or network output representing a saccade is similarly population coded, then the projections from the input to the target layer define a probability distribution over the output layer.
We can simplify the representation further by discretizing the space and considering the inputs and outputs as binned ranges, i.e., as a bivariate histogram.

Population codes lend to a natural interpretation of neural activity in terms of the primary operations used in probabilistic reasoning: marginalization and inference.
Because a receiving layer integrates over input from a sending layer, co-activation of multiple inputs to a single set of outputs is equivalent to marginalization (also known as mixture) over the conditional distributions produced by each neuron.
Each individual input may be further regarded as a histogram that marginalizes over the probability mass values represented by its projections.
Evidence accumulation for inferences may be best understood in terms of the cumulative change in synapses over long-term potentiation (LTP) and long-term depression (LTD). 
In statistical terms, synaptic plasticity itself serves as the multiplication of likelihoods, including priors, leading the posterior distribution encoded over synaptic connections. 
The prior distribution, under this interpretation, is just the previous state of the synapses, or the cumulative result of learning up that point.
\cite{aitchison2021synaptic} discuss the relationship between evidence and Bayesian inference as it pertains to synapses.
Their analysis is directly related to the property of maximum likelihood estimation in which the second partial derivatives of the likelihood function with respect to the parameters computes the variances and covariances of those parameters at the solution, and therefore the first derivatives, or step sizes, are a function of underlying parameter uncertainty \citep{lehmann2006theory}.
Inference, more broadly, can validly refer to both this abstracted relationship between evidence and learned synaptic states and the active use of those synaptic states by the network to process ambiguous stimuli.

\begin{figure}[!ht]
    \centering    
    \includegraphics[width=.65\linewidth]{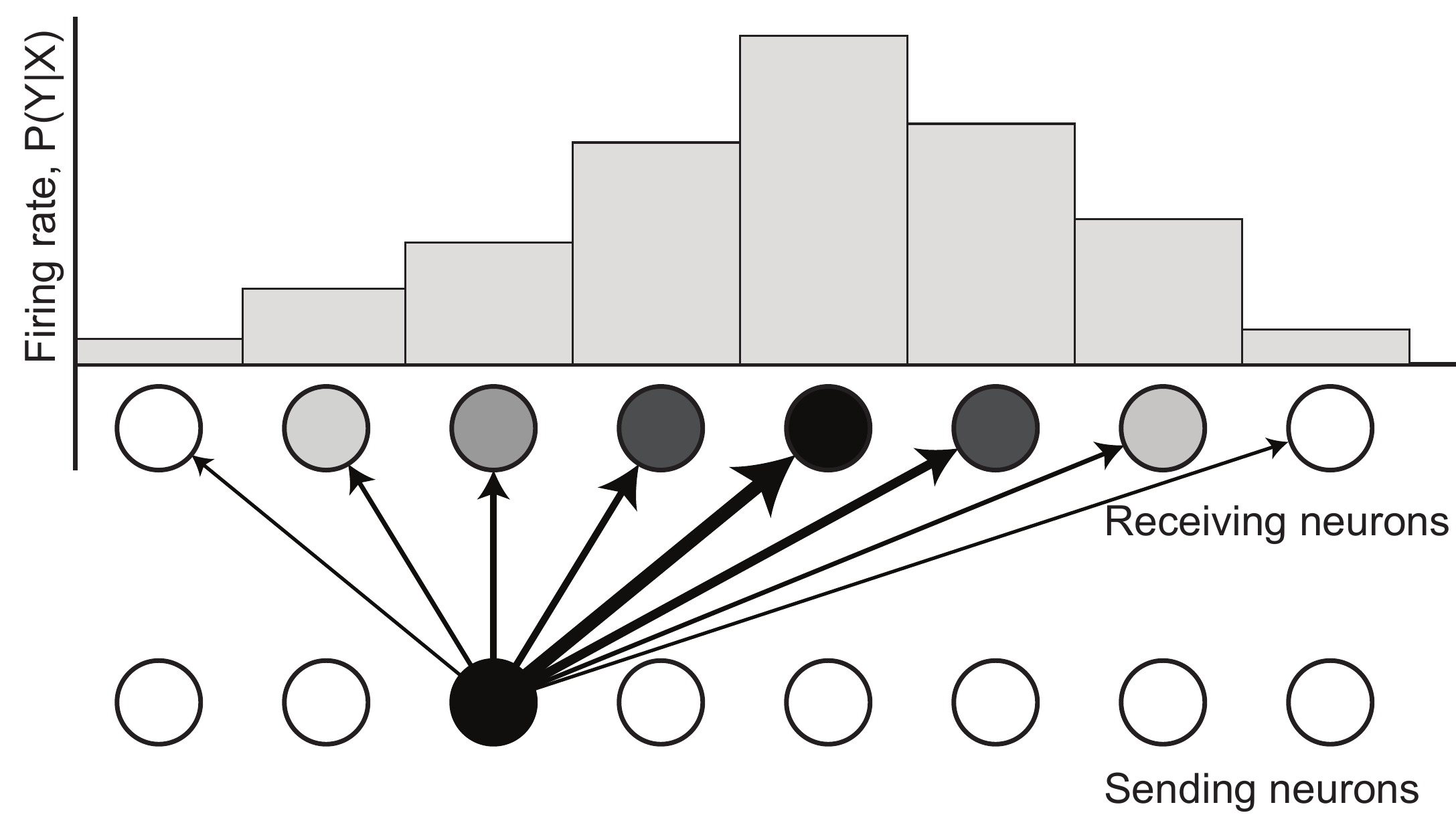}
    \caption{Neural activations may be interpreted as probability mass functions on a distribution that marginalizes over the receiving layer. Probabilities are ultimately encoded in the projections from the sending layer.}
    \label{fig:neuron2hist}
\end{figure}

\paragraph{Population codes as probability mass}
By combining population codes with sampling, the brain can represent probability distributions over two dimensions: space and time.  
Across space, the overall distributed pattern of neural activity across the population code can represent variance in terms of the breadth of neurons active: if only a single neuron is strongly active, the corresponding value is represented with low variance (high certainty); and if many neurons are weakly active, that corresponds to a high variance (low certainty).  
The nature of the learned weights clearly will drive the shape of this neural activity distribution.
Formally, one only needs to normalize the neural activations by the total activity to have a well-defined probability distribution (Figure \ref{fig:neuron2hist}).

Furthermore, with a few additional assumptions, the propagation of a population of distributed neural activity from one layer to the next will appropriately propagate the relative level of uncertainty into subsequent stages of processing.  
For example, if the size of a particular dog is uncertain, then many neurons representing different possible sizes will be active.
The widespread activity will cause further widespread activation in a subsequent layer tasked with deciding best size for a dog house, effectively \emph{convolving} uncertainty estimates computed at the next level with the uncertainty present in the lower level.

It has been postulated that neural activations map directly to likelihoods or log-likelihoods, a concept known as probabilistic population codes \citep{pouget2013probabilistic, pouget2003inference, beck2012complex, ma2006bayesian}.
Though several interpretations exist, the common theme is that each provides a way to decode neural activations in terms of statistical formalisms and operations. 
\cite{zemel1998probabilistic} examined three possible schemes for decoding neural activity as probabilistic inference.
In the first, distributions are implied by the likelihood function of a Poisson rate parameter conditional on the rate of the input neuron.
This approach is limited in the range of variances that can be represented, and functions more literally as an analysis of signal fidelity among neurons rather than a language of probability among them.
Furthermore, it only applies to inputs that are point values.
To remedy these issues, they extend the approach such that the Poisson rate parameter is the inner product of the tuning curves with an arbitrary input function, effectively allowing simultaneity among many inputs.
A somewhat different approach considered by the same authors is to represent distributions as a weighted sum of tuning curves, a form of kernel density estimation, but that approach entails implausible lower bounds on the precision and constraints between precision and location.
Generally, these approaches focus on inference in the space of neural activations, rather than through the long-term course of learning.

The purely spatially distributed population coding case represents a kind of static, ``omniscient'' perspective on probabilistic processing, where an entire distribution is represented at once.  At the other extreme is a purely time-based encoding of the distribution, which is synonymous with representing discrete \emph{samples} from the overall distribution \citep{HoyerHyvarinen2003, fiser2010statistically, orban2016neural}. For example, at the relevant sampling timescale, the population code could engage in a strong inhibitory competition to select a single neuron, with the statistics of this competitive process following the relevant underlying probability distribution (i.e., higher probability neurons are activated more frequently). This is the idea behind the widely-used SoftMax distribution, for example, where the likelihood of activating each neuron in a layer is normalized by the sum over all neural activations in that same layer. Theories of such probabilistic sampling over time underlie proposals for how noisy neural activity might implement non-synaptic-failure-based inference algorithms \citep{Buesing11, Hennequin14_fastsampling, SavinDeneve14}, have been used to explain stochasticity in synaptic weights \citep{Kappel15, aitchison2021synaptic}, and can explain aspects of response variability in sensory cortices \citep{orban2016neural, Echeveste20_corticallike}.

\paragraph{Synaptic failure as dropout sampling}
It is in this time-based sampling case that synaptic dropout can play a critical role, acting as the source of stochastic variability that drives the sampling process.  
As noted above, there is extensive evidence that synapses in the mammalian neocortex exhibit high rates of communication failure, specifically the failure of presynaptic vesicles to release neurotransmitters \citep{allen1994evaluation}.
In vivo, release probabilities vary widely as a result of maturational differences, extracellular calcium \citep{dodge1967co}, inhibition by ambient neurotransmitters, synapse type, and postsynaptic cell type, with most falling well below 50\% \citep{borst2010low, branco2009probability}.
\cite{huang1997estimating} estimated that two-thirds of synapses transmit subsequent to fewer than 17-33\% of presynaptic action potentials.

Interestingly, a purely temporal sampling scheme does not directly convey the uncertainty associated with a particular sample to subsequent processing stages, in contrast to the space-based representation.  
Repeated samples over time are required to gather a sense of the relative variance in neural activities.
For this reason, it may be particularly powerful to combine some of both the space and time representations of a distribution, representing variance across a relatively sparse population of neural activity, while selecting a sample from the larger overall distribution.  
It is difficult to analyze such a hybrid scheme, but given our ability to apply the same framework to both the purely spatial and temporal schemes separately, there is no obvious reason why it should not work in practice.

\paragraph{Lateral inhibition}
In addition to synaptic dropout, sampling in the brain is going to be affected by inhibitory connections among the receiving neurons.
The effect of these connections, called \emph{lateral inhibition}, can produce contrast and winner-take-all dynamics. Without sampling, a complex, multi-modal distribution subject to lateral inhibition may disproportionately express only the mode represented by the few strongest synapses. 
If synapses fail randomly, then competition is randomly reduced and all neurons constituting the distribution are given a chance to fire over the course of repeated samples.
In fact, the network cannot be said to sample from an encoded distribution unless each neuron's ultimate chance of out-competing the rest of the sample is exactly the probability encoded by its incoming synapses.
If transmission probabilities are simply uniform or proportional to the synaptic weights, then the likelihood of sampling the tails of the distribution, encoded by the smallest weights, vanishes as the size of the network grows.
For the theory we have outlined to be possible, the probability that a synapse fails must be systematically related to its efficacy, but no functional mappings have been found previously.
This final observation motivates the central aims of this paper: to discover such a mapping and demonstrate biologically plausible sampling of all the distributional components relevant for Bayesian inference.

In summary, we make the following theoretical proposals: (1) Probability distributions are encoded in the sending projections and manifested in the receiving layer as population codes; (2) Synaptic plasticity constitutes the accumulation of evidence and inference between layers; (3) Integration over a sending layer constitutes marginalization over conditional distributions in the receiving layer; (4) Synaptic failure serves as a dropout-type sampling algorithm; (5) Transmission probabilities adapt in tandem with weights during learning to sample from distributions that account for both observed and unobserved variance, i.e., that of sensed data and model parameters (in neurons, synaptic efficacy itself).
All together, these propositions allow for the complete realization of approximate Bayesian inference in the brain.

In what follows, we describe a formal framework based on the histogram-like interpretation of population codes that enables both spatial and temporal representations of arbitrary probability distributions.
We formulate a sampling scheme that consists of iterations of synaptic dropout followed by lateral inhibition.
Within this framework, we find a mapping from synaptic weights to transmission probabilities that enables sampling from encoded distributions that include both observed and epistemic variance components.
From our result, we then derive and demonstrate a locally-computable learning rule for synaptic transmission probabilities that is consistent with recent biological findings.

\section{Theoretical model}
Our theoretical model incorporates several constraints and design considerations motivated by an attempt to capture some basic properties of biological networks: (1) Continuous inputs and outputs are represented in the space of neural rate codes by way of neural tuning curves; (2) Both firing rates and synaptic weights exhibit physical lower and upper bounds, or saturation points; (3) Weights representing the encoded distribution must be non-negative, consisting of  excitatory connections; 
(4) Neuron-to-neuron operations in learning and inference must be local, using only the available state of the network at a given time. Additionally, it is important to consider the effects of lateral inhibition and the resulting in winner-take-all dynamics among receiving layers, which is common throughout the cortex.

\subsection{Evidence accumulation}
To start, a neural network could be considered Bayesian if the distribution of free parameters, i.e., the weights or synapses, $P(\mathbf{W})$, updates according to Bayes' theorem as new data are introduced.
It is not the goal of this paper to show that this is true of synapses, but it is a prerequisite for epistemic uncertainty and the premise of ongoing speculation \citep[e.g.,][]{aitchison2021synaptic}.

Bayes' theorem for updating weights $\mathbf{W}$, given input activations $\mathbf{X}$ and output activations $\mathbf{Y}$ is given by:
\begin{align}
        P_{t+1}(\mathbf{W}|\mathbf{X}, \mathbf{Y}) &= \frac{P_t(\mathbf{X}, \mathbf{Y} | \mathbf{W})P_t(\mathbf{W})}{P_t(\mathbf{X, Y})},
\end{align}
The model is conditioned as $\mathbb{E}[\mathbf{Y}]=f(\mathbf{X}, \mathbf{W})$, where $f$ is most often a sigmoidal, reLU, or radial link function in artificial neural networks.
For simplicity, take $\mathbf{X,Y,W}\in[0,1]$, as both biological synaptic weights and neural firing rates are positive valued with an upper saturation boundary.
Under these constraints, $f$ may simply be the dot product $\mathbf{XW}^\intercal$.
The beta distribution then gives us a mean field approximation of $P(\mathbf{W})$ that respects the [0,1] interval:
\begin{align}
            \text{Beta}(p; \alpha, \beta) &= \frac{p^{\alpha-1}(1-p)^{\beta-1}}{\int_0^1 x^{\alpha-1}(1-x)^{\beta-1}dx}, \\
            P(\mathbf{W}) &\sim \text{Beta}(\mathbf{A}+1,\mathbf{B}+1), \label{eq:beta}
\end{align}
such that weights are distributed according to the cumulative evidence for and against each association represented by the cells of matrices $\mathbf{A}$ and $\mathbf{B}$, respectively:
\begin{align}
    \mathbf{A} &= \mathbf{A_0}+\lambda \mathbf{X^\intercal Y},\quad
    \mathbf{B} = \mathbf{B_0}+\lambda \mathbf{X^\intercal (1-Y)}.
\end{align}
Zero-subscripted matrices represent the prior cumulative evidence, and $\lambda$ is a learning rate or weight assigned to the new evidence.
The matrices of evidence for and against each possible input-output association are based on the logic of long-term potentiation and depression, respectively.
$\mathbf{A}$ sums over the pairwise products of input and output activations, representing the extent to which inputs and outputs fired jointly, while the matrix of evidence against each association is given by $\mathbf{B}$, the extent to which outputs did not fire following input activations.

The matrix of mean weight values is calculated as,
\begin{align}
    \mathbb{E}[P(\mathbf{W})] &= \frac{\mathbf{A}}{\mathbf{A}+\mathbf{B}}. \label{eq:betamean}   
\end{align}

Note that whereas the mean of $\mathbf{W}$ is simply the normalized evidence for association, the precision of $\mathbf{W}$ scales with the non-normalized, cumulative values in $\mathbf{A}$ and $\mathbf{B}$ as
\begin{align}
\text{var}(\mathbf{W}) &=\frac{\mathbf{A\odot B}}{(\mathbf{A+B})^2(\mathbf{A+B}+1).}\label{eq:betavar}
\end{align}

The above equations define a variational model in which the solution and approximate distribution of the parameters are analytically computed from the data.
Weights may be sampled from Equation \ref{eq:beta} and used to produce a posterior distribution of the expected values for the output layer given any particular input vector.
Variational approximations of $P(\mathbf{W})$ do not, however, provide the complete conditional distribution of output activations.
Rather, they give the distribution of the \emph{mean} activation conditional on the parameters, which will depend on $\lambda$ and the sample size.

This beta-distribution based model conveniently serves our purposes in three ways: (1) It simplifies our task of abstracting from biological networks and preserving their domain constraints; (2) It allows learning to be directly derived from the rules of long-term potentiation and depression; (3) It provides a quasi-analytic expectation for parameter variance that we will later use as a point of reference for the biologically-motivated alternative of weight dropout.
Notably, variational, mean-field approximations such as this do \emph{not} give the true, ``full Bayes'' posterior of the weights, which is implausible in the brain and intractable in artificial networks.

\subsection{Epistemic uncertainty}
We can use the mean-field beta distribution above, namely Equation \ref{eq:betavar}, to derive transmission probabilities that produce roughly the same epistemic uncertainty.
That is, we want to transfer the properties of the beta model to an equivalent mean-field Bernoulli approximation.
With dropout, the actual distribution of the parameters is a scaled Bernoulli:
\begin{align}
    w_i &= \alpha_i z_i,\quad z_i \in \{0,1\},\quad P(z_i=1) = \phi_i.
\end{align}
To produce an analogous distribution to the beta model in terms of dropout, we must equate the variances and solve for transmission probability $\phi_i$.
Note that the mean value of the weight given dropout is $w_i\phi_i$.
If we are equating the mean and variance to the beta model, we must use $\hat w_i = w_i / \phi_i$.
With the matrix of transmission probabilities, $\Phi$, and denoting the element-wise Hadamard product with $\odot$,
\begin{align}
    \mathbf{\hat W}^2 \odot \mathbf{\Phi \odot (1-\Phi)} &= \frac{\mathbf{A\odot B}}{(\mathbf{A+B})^2(\mathbf{A+B}+1).}. \nonumber \\
    \left(\frac{\mathbf{A}}{\mathbf{\Phi\odot(A+B)}}\right)^2 \mathbf{\Phi\odot(1-\Phi)} &= \frac{\mathbf{A\odot B}}{(\mathbf{A+B})^2(\mathbf{A+B}+1)}.
\end{align}

Solving for $\mathbf{\Phi}$ when $\mathbf{\Phi, A,B}>0$, gives us
\begin{align}
    \mathbf{\Phi} &= \frac{\sqrt{(-\mathbf{A}^2-\mathbf{AB-A})^2 + 4\mathbf{B(A}^2+\mathbf{AB+A})} - \mathbf{A}^2-\mathbf{AB-A} }{2\mathbf{B}} \label{eq:beta2dropoutvar}
\end{align}

We can then perform dropout sampling of the posterior distribution over unobserved values, or epistemic uncertainty, by generating binary mask matrices from $\mathbf{\Phi}$ that then element-wise multiply by the weight matrix.
For minimal bias, the weight matrix should be element-wise divided over $\mathbf{\Phi}$, such that the mean value of each weight is $\mathbf{W}$. 

Simplified approximations of the above function are possible.
For instance, the special case of $\mathbf{A}=\mathbf{B}$ is close to $\Phi=1-\frac{1}{\mathbf{A+B+3}}$.
Whereas these results produce the kind of epistemic uncertainty familiar in classical statistics, i.e., sampling error of the parameters, in the brain, epistemic uncertainty may be modulated in response to numerous contextual factors, making the above result only a special case.


\subsection{Transmission probability from synaptic weight}
Unlike epistemic uncertainty, there is no clear precedent for deriving the observed data distribution using variational modeling principles alone.
Here we will consider a first-principles approach using the abstract model defined so far.
If we take $\mathbf{X}, \mathbf{Y} \in \{0, 1\}$, i.e., a discrete activation space, and we consider the special case of $m=1$ input neuron $x$, then the $1\times n$ weight vector $\mathbf{w}$ is exactly a vector of observed frequencies resulting from the conditional distribution $P(\mathbf{Y}|x=1)$.
Activation of multiple inputs, each representing a unique conditional distribution, therefore corresponds to a marginalization (i.e, mixture) of each in the output layer.
We will use this fact to determine a rule for mapping weight values to transmission probabilities such that the frequencies of output samples produced by a dropout algorithm are their learned conditional probabilities.

In biological networks, we propose that sampling of the conditional distribution involves two mechanisms.
First, random synaptic failure results in only a subset of active outputs.
Then, the most active output neuron among the subset suppresses the others by lateral inhibition.
This process is approximated by an artificial sampling scheme in which weights are randomly set to zero, and then the maximum output activation is chosen from the resulting subset.

\begin{figure}[!ht]
    \centering
    \includegraphics[width=\linewidth]{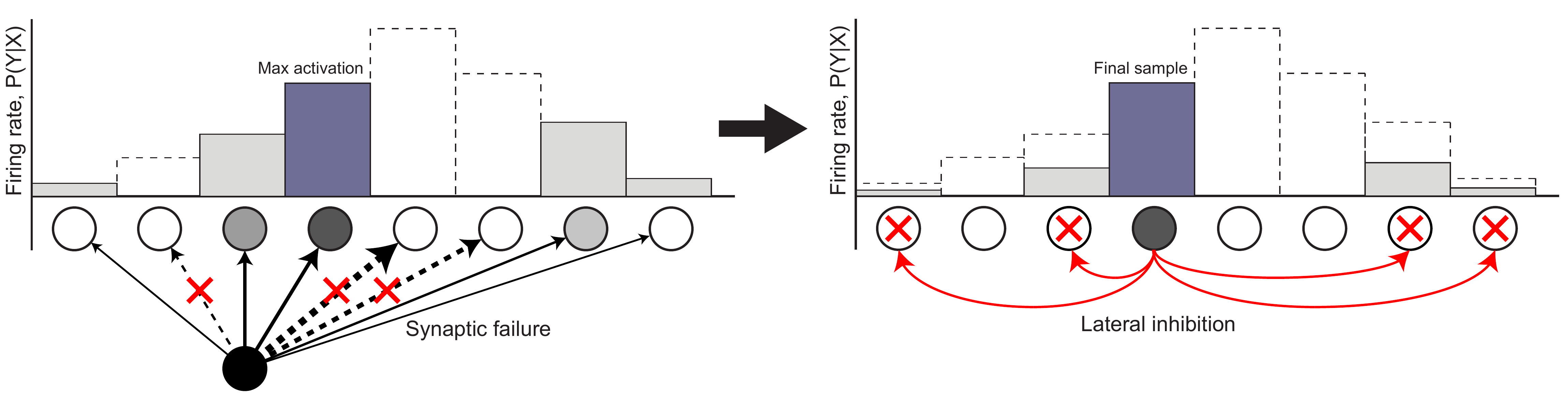}
    \caption{A sample of the encoded posterior distribution is drawn through synaptic failure followed by lateral inhibition. A subset of activations in the receiving layer first results from the random failure of several sending synapses. Then, lateral connections from the most active receiving neuron inhibit the competing receiving neurons, resulting in a single, random realization from the complete distribution.}
    \label{fig:samplediagram}
\end{figure}
To find a mapping of weights to transmission probabilities for this subset-max scheme, consider the encoded histogram in descending order, where $i\in 1,...,n$, and $w_1 > w_i > w_n$, and let $q_i$ be the final transmission probability for weight $i$.
Let $p_i$ represent the encoded probability of sampling output activation $i$, defined as the normalized synaptic weight $w_i/\sum_{i=1}^n w_i$.
Because we are taking only the maximum weight from each subset, the probability of sampling from the largest of all weights is just its normalized value: $q_i=p_i$.
For each successive weight of rank $i>1$, the encoded probability must be equal to the probability that weight $i$ transmits and all larger weights fail:

\begin{align}
    p_i &= q_i\prod_{j=0}^{i-1}(1-q_j),
\end{align}

where $q_0=0$.
Solving recursively for $q$ in terms of $p$, we find 

\begin{align}
    p_1 &= q_1, \nonumber\\
    p_2 &= q_2(1-p_1), \nonumber\\
    p_3 &= q_3(1-p_1)(1-q_2)\nonumber\\ 
        &=q_3(1-p_1)(1-\frac{p_2}{1-p_1}) \nonumber \\
        &=q_3(1-p_1-p_2),\nonumber\\
    p_4 &= q_4(1-p_1)(1-q_2)(1-q_3) \nonumber \\
        &=q_4(1-p_1)(1-\frac{p_2}{1-p_1})(1-\frac{p_3}{1-p_1-p_2}),\nonumber\\
        &=q_4(1-p_1-p_2-p_3), 
\end{align}
and so on for all $p_i$, giving us
\begin{align}
    p_i &= q_i(1-\sum_{j=0}^{i-1}p_j), 
\end{align}
which is the probability that weight $i$ does not fail and no larger weights are ultimately sampled.
A complete proof of this result by induction is given in the appendix.
The transmission probability for weight $i$ is therefore
\begin{align}
    q_i &= \frac{p_i}{1-\sum_{j=0}^{i-1}p_j}  \nonumber \\
    &= \frac{w_i}{(\sum_i w_i)(1-\sum_{j=0}^{i-1}\frac{w_j}{\sum_i w_i})} \nonumber \\
    &=\frac{w_i}{\sum_i w_i - \sum_{j=0}^{i-1}w_j }\nonumber \\
    &= \frac{w_i}{\sum_{j=i}^n w_j}.  \label{eq:mapping}
\end{align}

For $m>1$ input neurons sharing an output layer, we conjecture that no general solution exists such that transmission probabilities are held constant across combinations of active inputs.
The above single-input mapping simplifies our derivation because the output activation $y_i = w_i$, so $p_i$ is taken to be $w_i$ normalized over the sole vector of input weights to which it belongs.
For multiple inputs vectors, $p_i = f(\mathbf{W}^\intercal \mathbf{x})$, and so normalization of a particular weight is dependent on $\mathbf{x}$ and no longer constant.
It may be possible that separate solutions exist for every possible combination of input activations over $\mathbf{x}$, in which case transmission probabilities are dynamic with respect to input activations.

When multiple inputs are active, probabilities must account for the possibility of being out-competed by projections from other inputs.
The simplest approximation is to compute Equation \ref{eq:mapping} for each vector of weights, then divide all probabilities by the total number of active input neurons.
In a model of continuous data with tuning curves such that activations are continuous between zero and one, the denominator may be the sum of the input vector.
Numerically and biologically, the same problem may be solved by simultaneous, local learning applied across the set of all relevant weights, as we will see later.  

\paragraph{Combined uncertainty}
Altogether, samples from the posterior distribution over the receiving layer should vary according to both the observed data and unobserved values in the model, i.e., both aleatoric and epistemic sources of uncertainty. 
As we have now derived particular probability functions for both, the functions can be multiplied, i.e., $\mathbf{\bar Q} = \mathbf{\Phi \odot \mathbf{Q}}$, where $\mathbf{\bar Q}$, to produce the final dropout probability matrix used to generate a mask over $\mathbf{W}$.

\subsection{Local, biologically plausible learning rules}
Equation \ref{eq:mapping} compels us to search for a biologically plausible algorithm by which synaptic release probabilities are updated.
The algorithm must not involve global operations and draw only from the present state of the network.
We propose two algorithms by which probabilities converge to their analytically shown solution over the course of posterior sampling.
To start, a generic recurrence equation for learning of transmission probability $i$ at sampling iteration $t$ is
\begin{align}
    \hat q_{i,t} &= \hat q_{i, t-1} + \gamma (q^*_i - \hat q_{i,t-1}), \label{eq:learningRule_form}
\end{align}
where $q^*_i$ is the target value of the learned transmission probability $\hat q_i$.
With the previous rank-order notation, $w_1 > w_i > w_n$, the set $\mathbf{s}_t$ of non-zero input weights after dropout allows for a biased approximation to Equation \ref{eq:mapping}, which we can use as target $\hat q^*_i$ of our learning rule: 
\begin{align}
    \hat q^*_i = \frac{w_i}{\sum_{j \in \mathbf{s}_t} w_j}.
\end{align}
Critically, $w_i$ is the maximum weight in $\mathbf{s}_t$, and so only the maximum is updated per sample.

The form of this target is possible to instantiate biologically if vesicle release probability is modulated by an extracellular agent that is available in limited supply during each iteration of sampling.
If synapses compete to take up the agent at rates that are a function of their efficacy, then the amount taken up by the largest synapse corresponds to the normalized value above.

The approximate target given above differs from the analytic target in that the denominator sums over only a subset of $w_i...w_n$, and so estimates will be upwardly biased.
There may be many ways to correct for that bias with varying degrees of biological plausibility and effectiveness.

\paragraph{Fixed adjustment}
The simplest method of correcting bias may be to subtract a fixed value from the target:
\begin{align}
    \hat q^*_{i,t} = \frac{w_i}{\sum_{j \in \mathbf{s}_t} w_j}-c.
\end{align}
This may correct for overall probability but result in inaccurate scaling of the relative differences among $\hat q_i$.

\paragraph{Scale adjustment}
As $|\mathbf{s}_t|\leq (n-i+1)$ and $\mathbf{s}_t \subset \{i...n\}$, $\hat q^*_i$ is larger than $q_i$ on average by a factor of $\frac{n-i+1}{|\mathbf{s}_t|}.$
We could multiply $\hat q_i$ by the inverse of this factor to correct for average bias across $\hat q_i$.
\begin{align}
    \hat q^*_{i,t} = \frac{|\mathbf{s}_t|w_i}{(n-i+1) \sum_{j \in \mathbf{s}_t} w_j}
\end{align}
To the extent that the weights are not uniform, this will over-correct relative differences among $\hat q_i$, resulting in shrinkage in the tails of the distribution.

\paragraph{Exponential adjustment}
Raising the approximate target to an exponent, $\psi$ results in shrinkage of the target that preserves the relative differences among transmission probabilities.
\begin{align}
    \hat q^*_{i,t} = \left(\frac{w_i}{\sum_{j \in \mathbf{s}_t} w_j}\right)^\psi.
\end{align}

There may be many other possible learning rules that achieve adequate approximation of $q_i$, with varying degrees of biological plausibility.
We consider only these few to show that it is possible.
The form of the target that we have outlined here, $\hat q^*_i$, makes Equation  \ref{eq:learningRule_form} a nonlinear stochastic recurrence equation, which has no explicit form for its equilibrium state and is difficult to analyze.
For further details, see Appendix B.

\paragraph{Algorithms}
In the first algorithm, the update is performed once per sample, affecting only the largest member:
\begin{lstlisting}
for i iterations:
  Generate Mask from Bernoulli(q)
  Subset = Mask*Weight
  m = max weight index from Subset
  g = Subset/Sum(Subset)
  q[m] = q[m] - LR(q[m] - g[m]^psi) 
\end{lstlisting}

In the second algorithm, the update is performed in descending rank order on all members of the subset per sample.
Each maximum is masked after its update, producing a smaller subset of size $|\mathbf{s}_t|-1$ and an update to the next largest member:

\clearpage
\begin{lstlisting}
for i iterations:
  Generate Mask from Bernoulli(q)
  Subset = Mask*Weights
  while Subset contains non-zero values:
    m = max weight index from Subset
    g = Subset/Sum(Subset)
    q[m] = q[m] - LR(q[m] - g[m]^psi) 
    Subset[m] = 0
\end{lstlisting}

This second algorithm converges to a reasonable approximation of $\mathbf{q}$ in far fewer samples than the first, but requires a more complex sequence of events within-sample.
Either algorithm may involve a gradually decreasing learning rate (\texttt{LR}), and both may include boundaries to fix $q$ between zero and one.
A small, non-zero lower bound is preferable, as a true zero will eliminate the synapse from all future samples, and consequently, any further learning.
Whereas analytically obtained dropout weights must be down-scaled to account for multiple active inputs, the above algorithms may be applied to a pooled set of multiple weight vectors so that no adjustment is necessary.

\paragraph{Learning rule simulations}
To compare learning rules, a weight vector representing an encoded bimodal distribution was generated.
Each learning rule was used to train a vector of transmission probabilities, that was then used to draw random samples from the weight vector.
Sampling was performed by first applying the dropout mask, then taking the maximum of the remaining weights.
Transmission probability start values were set to $\hat q_{i,0} \sim N(0.3, 0.1)$, and each learning rule was applied with Algorithm 2 for 10,000 iterations at learning rate $\gamma=0.0025$.

\begin{figure}[!ht]
    \centering
    \includegraphics[width=\linewidth]{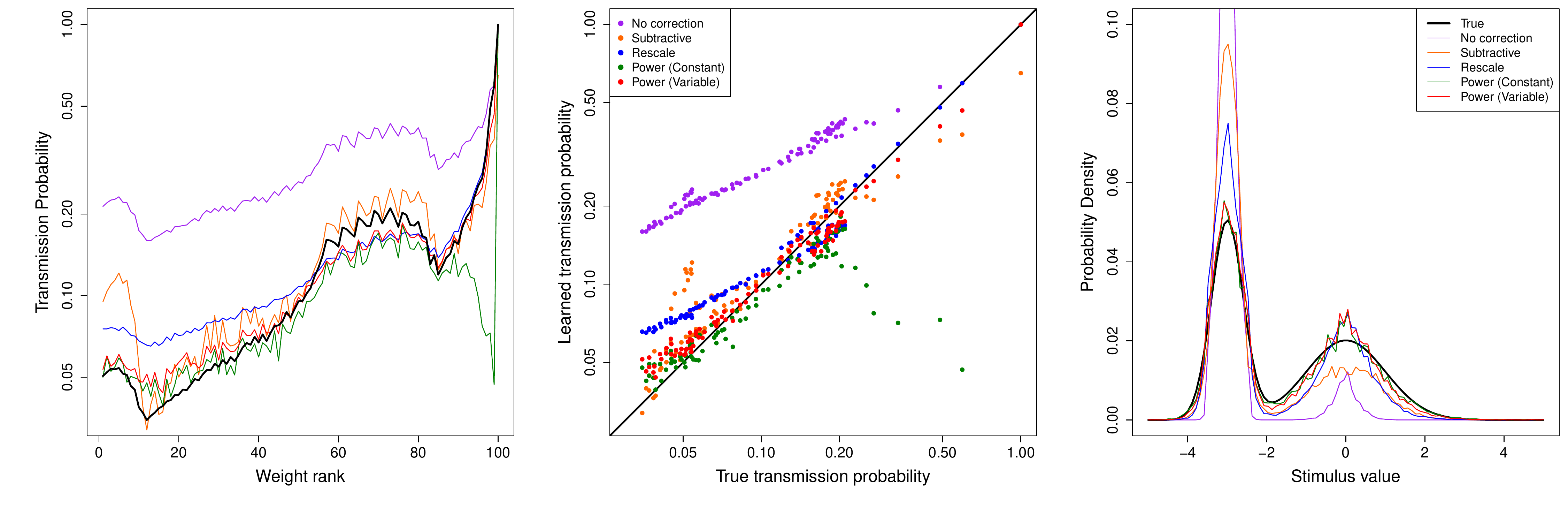}
    \caption{Comparison of learning rules with different bias adjustments to the target. Subtractive: a constant (0.35) is subtracted from $\hat q^*_{i,t}$. Rescale: $\hat q^*_{i,t}$ is rescaled by  $|\mathbf{s}_t|/(n-i+1)$. Power (Constant): $\hat q^*_{i,t}$ is raised to a fixed exponent ($\psi=7$). Power (Variable): $\hat q^*_{i,t}$ is raised to a varying exponent ($\psi=(n-i)\hat q_{i,t-1}+1$).} 
    \label{fig:learningRules}
\end{figure}

Figure \ref{fig:learningRules} compares the results of five possible learning rules under Algorithm 2, each with a different strategy for adjusting the target bias. 
In the left panel, analytic and learned transmission probabilities are plotted against the rank of their associated weights, in order for largest to smallest.
In the center panel, learned probabilities are compared against the analytic values.
In the right panel, the true distribution is compared to each learned result.

First, it can be seen in the left and center panels that using the unadjusted target for learning results in probabilities (in purple) that retain the form of the analytic probabilities (black) but differ in both scale and offset.
Subtracting a small value from the target preserves but exaggerates features of the distribution (orange). 
Rescaling the target by the given subset size over its max value results in transmission probabilities biased toward their mean value (blue).
Raising the target to a constant exponent $\psi$ results in substantially less biased results (green) in all but the smallest weights, which would otherwise be assigned the highest transmission probability.
Using a variable exponent $\psi_t=(n-i)\hat q_{i,t-1} +1$ (red) corrects for all bias, including the smallest weights.

The panel on the right shows that subtractive and scalar adjustments to the learning target do not adequately suppress the largest activations, resulting in under-represented tails.
Conversely, both exponent based adjustments produce very close approximations to the full distribution.
Bias in the smallest weights has little effect on the final distribution, so the gains of using a variable exponent over a fixed exponent are negligible.
Interestingly, the fixed exponent adjustment has particular support from \cite{dodge1967co}, who found that the postsynaptic potential of muscle fibers was related to the fraction of an inferred receptor type bound by calcium ions raised to the fourth power.
They infer that this relationship results from the cooperative action of four calcium ions per vesicle release.

\subsection{Network Simulations}

Monte Carlo simulations were used to look at the average tendency of dropout to sample both the conditional distribution of the output data (i.e., Equation \ref{eq:mapping}) and epistemic uncertainty resulting from the parameter distributions (Equation \ref{eq:beta2dropoutvar}) in the model described previously.
Weight samples from the beta distribution were used as a semi-analytic point of comparison for the latter.
Analytic transmission probabilities and the local, iterative learning methods were compared.

\paragraph{Data}
Two data-generating models were used.
The first involved mapping discrete input values to output distributions with increasing variance, i.e., heteroskedasticity, to test the accuracy of estimated residual variances by the dropout model.
The second involved both heteroskedasticity and bimodality over a continuous range to examine the overall versatility of the model.

The first input and output data were generated from continuous normal distributions as $y\sim \mathcal N(0, 0.2+0.2(x+4)), x\in \{-4, -2, 0, 2, 4\}$, with $N=4,000$ rows of data total, with an equal number per value of $x, N_x = 800$.

The second model was a mixture distribution with $N=4,000$ rows of data total:
\begin{align}
y&\sim p(x)f_1(x) + [1-p(x)]f_2(x), \quad x\in [-4, 4] \nonumber \\ 
p(x)&=\text{logit}^{-1}(x/2),\nonumber  \\ 
f_1(x)&= \mathcal{N}(-2, 0.2),\nonumber  \\
f_2(x)&= \mathcal{N}(x/4, 0.2 + 0.0625(x+4)).   
\end{align}
In this scenario, one distribution has a fixed mean and standard deviation, but with a declining density, while the other has a positive linear trend in the mean, variance, and density.

For all simulations, $x$ and $y$ were coordinate transformed to multivariate tuning curve activation matrices $\mathbf{X}$ and $\mathbf{Y}$ using a kernel of 100 Gaussian curves, equally spaced in $[-6, 6]$ with $\sigma=0.05$.
To transform random posterior samples $\mathbf{\hat Y}=f[(\mathbf{M \odot W})^\intercal \mathbf{X}]$, where $\mathbf{M}\sim \text{Bernoulli}(\mathbf{\bar Q})$, to real numbers $\hat y$, we defined a sequence $z\in [-6,6] \subset \mathbb{R}$, a Gaussian kernel matrix $\mathbf{H}$ defined over $z$, and used
\begin{align}
     \hat y_i &= \underset{z_i}{\text{arg max}}\, \mathbf{H Y_i^\intercal }.
\end{align}

\paragraph{Models}
Learning network weights involved summing the outer products of $\mathbf{X,Y}$ as described previously.
The initial values, i.e., priors, for the evidence matrices were generated as $\mathbf{A_0} \sim \mathcal U(0.025, 0.026)$ and $\mathbf{B_0} \sim \mathcal U(0.100, 0.101)$, and the learning rate was set to 1.
1,000 samples per $x$ value were generated, and 200 repetitions of the simulation were run to produce distributions of each estimated variance component.

The beta distribution model was used as a reference to produce posterior samples from the distributions of weights, representing epistemic uncertainty, where $\mathbf{W}\sim\text{Beta}(\mathbf{A}+1,\mathbf{B}+1)$.

The primary model of interest used the fixed matrix of weight means, $\mathbf{W}=\mathbb{E}[\text{Beta}(\mathbf{A}+1,\mathbf{B}+1)]$ (Equation \ref{eq:betamean}), and relied only on dropout for both epistemic and residual variance components.
For each posterior sample, this weight matrix was masked by probability matrix $\mathbf{\bar Q}=\Phi \odot \mathbf{Q}$, i.e., transmission probabilities combining both empirical (Equation \ref{eq:mapping}) and epistemic (Equation \ref{eq:beta2dropoutvar}) distributions.
The model was run with both locally learned probability matrix $\mathbf{\bar Q}$ using Algorithm 2 with 5,000 iterations and the analytically obtained probability matrix, i.e, rows computed according to Equation \ref{eq:mapping}.
The learning rule with fixed exponential bias adjustment was used, with $\psi=8$ and $\gamma=0.01$.

\begin{figure*}[!p]
\centering
    \includegraphics[width=\linewidth]{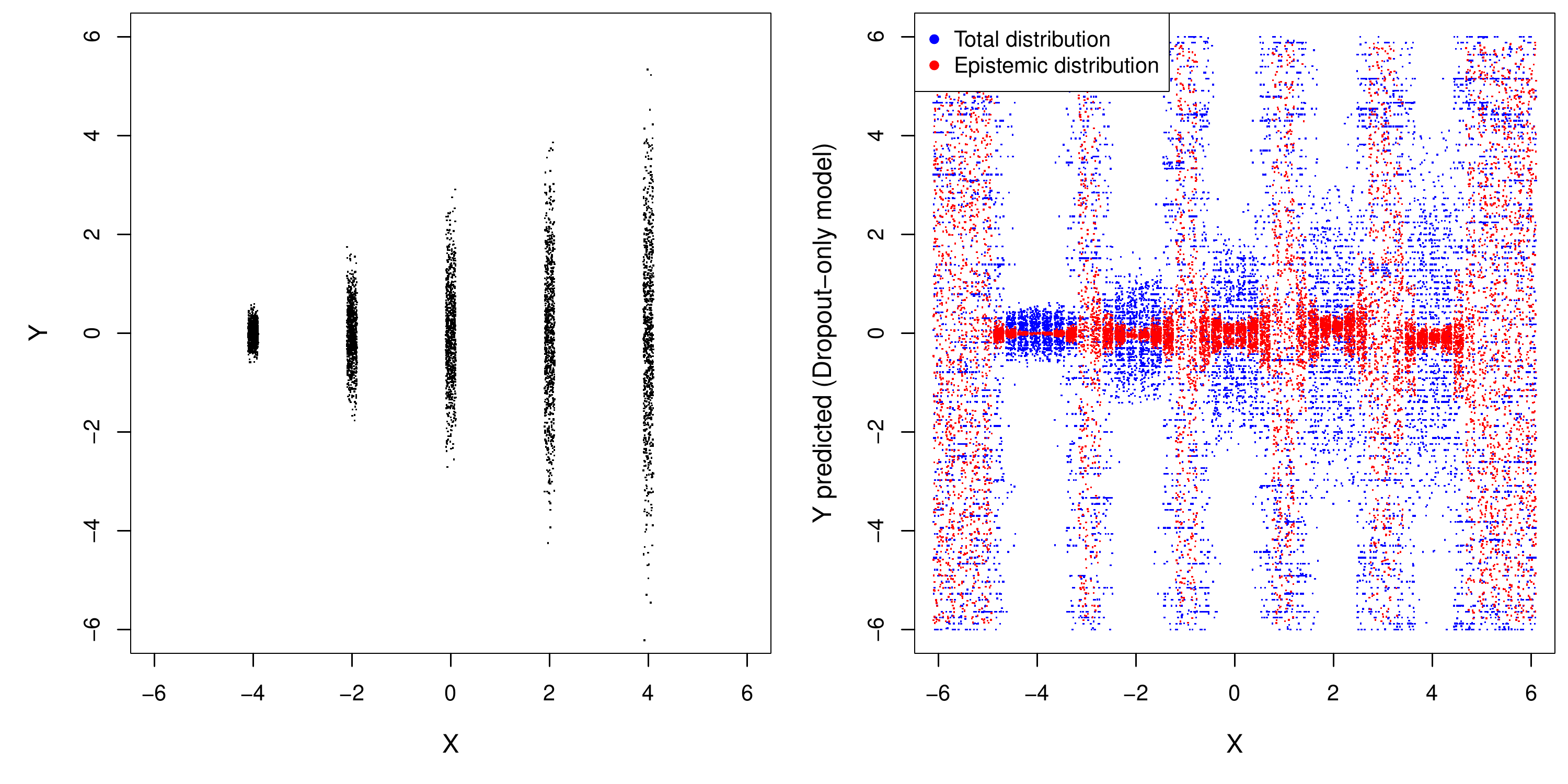}
    \includegraphics[width=\linewidth]{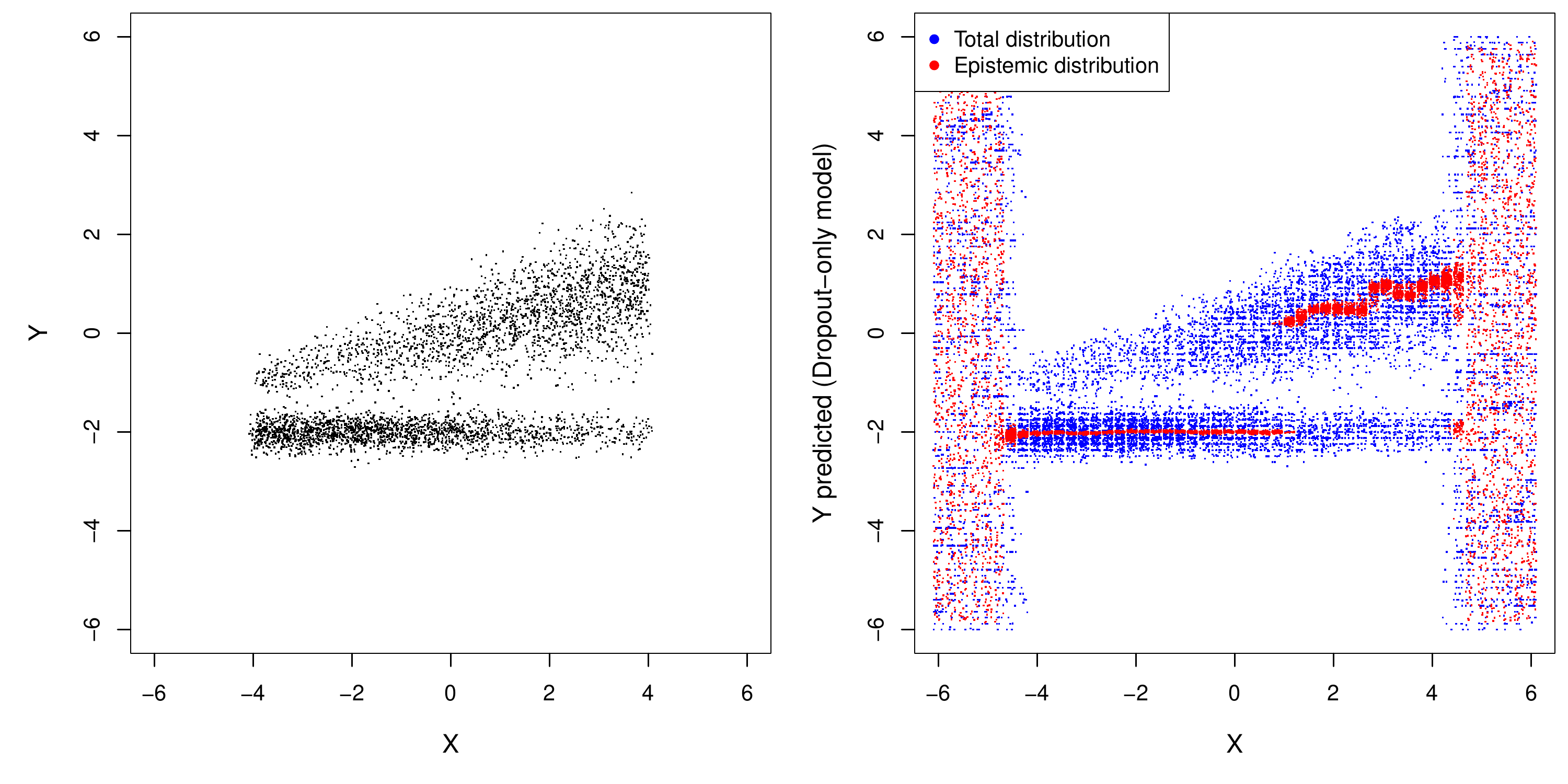}
    \caption{Simulated data versus dropout samples from the network. Top row: simple variance model using only five input values. Bottom: bimodal, heteroskedastic model.  Left column: One trial of simulated data from each model. Right column: Samples from the complete estimated posterior distribution (blue) and samples of the MAP only (red). X-axis jitter added to better show sample density.
    Epistemic variance is shown to vary with the amount of available data from which to learn each part of the input-output domain.
    Uniform priors are sampled where no data were available.}
    \label{fig:simdata1}
\end{figure*}

\paragraph{Results}
In our primary results, we visually and numerically compare simulated data to samples drawn from the posterior distribution encoded in the learned network weights.
If the network is accurately representing the complete Bayesian posterior, then the network samples should be distributed approximately according to the original data but with additional variance corresponding to epistemic uncertainty, i.e., inflation in regions with fewer data points.
Figure \ref{fig:simdata1} shows the simulated data (left) with the associated posterior samples over the full domain of $x\in[-6,6]$ (right).
Samples from the complete posterior distribution are shown in blue.
The additional epistemic uncertainty is overlaid in red as samples of the \textit{maximum a posteriori} (MAP).
The MAP was sampled by setting transmission probabilities to only Equation \ref{eq:beta2dropoutvar} such that the observed data distribution was excluded.
The network draws uniformly distributed samples where no data are available to inform input-output associations, such as between activated inputs values in the first simulation, and toward the edges of the input-output domains in both simulations.
The uniformity of the samples in these regions demonstrates maximal epistemic uncertainty.
Specifically, the network samples from the uniform priors of the network weights in the absence of any data to update those priors.
Conversely, the variance of MAP samples is smallest where the training data are most available, reflecting a high certainty in the most central or likely output value.
The gradations from maximal to minimal uncertainty visible around each input value in the first simulation reflects the weaker co-activation of tuning curves peripheral to each input value.

\begin{figure}[!ht]
    \centering
    \includegraphics[width=\linewidth]{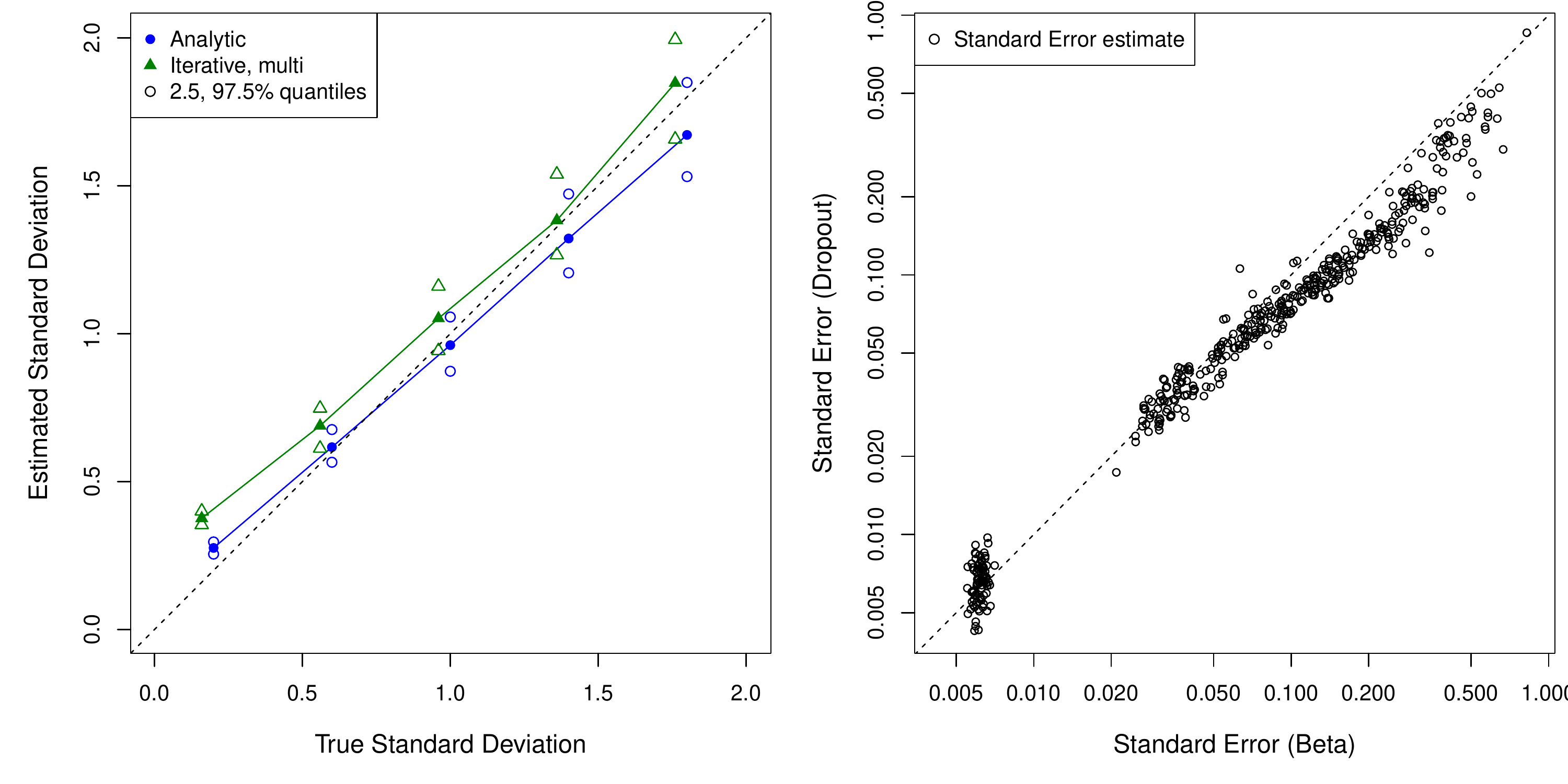}
    \caption{Left: Estimated standard deviations of posterior samples, excluding epistemic uncertainty, versus the true, data-generating standard deviations. Both analytic and iteratively learned transmission probabilities produce accurate samples of the observed distribution with minor bias.
    Right: Epistemic uncertainty comparisons as MAP standard error estimates obtained by sampling from the Beta model (x-axis) versus by dropout (y-axis).
    Open symbols represent 95\% quantiles.}
    \label{fig:simresults}
    \label{fig:weights}
\end{figure}

The first simulation was used to make numeric comparisons of the posterior distribution to the data-generating model.
The left panel of Figure \ref{fig:simresults} shows the average estimated standard deviations of output samples for each active input value.
These samples used only the analytic and learned probabilities representing the observed distribution, whereas epistemic uncertainty was excluded.
Results for both the analytic and iteratively learned transmission probabilities are plotted against their respective data-generating values.
Both methods closely approximate the true standard deviations with minor bias.
An overall upward bias in the local learning reflects imperfect convergence to the analytic probability function, which can be improved with more iterations and a lower learning rate.
The analytically derived transmission probabilities were less biased overall, but with slight downward bias in the highest variances, and slight upward bias in the lowest.
These biases may reflect either or both imperfect choices for normalizing over multiple inputs or an artifact from our chosen strategy for inverting the tuning curves to produce the final continuous samples.

The right pane of Figure \ref{fig:simresults} compares the standard error, i.e., standard deviation of MAP samples, from the transmission probabilities to estimates given by samples from the beta distribution model.
Here, too, results fell along the identity line, showing that dropout produces a variational model with comparable results to the beta distribution.
This result is expected because the epistemic adjustment to transmission probabilities was derived from the variance equation for the beta distribution.

\begin{figure}[!ht]
    \centering
    \includegraphics[width=\linewidth]{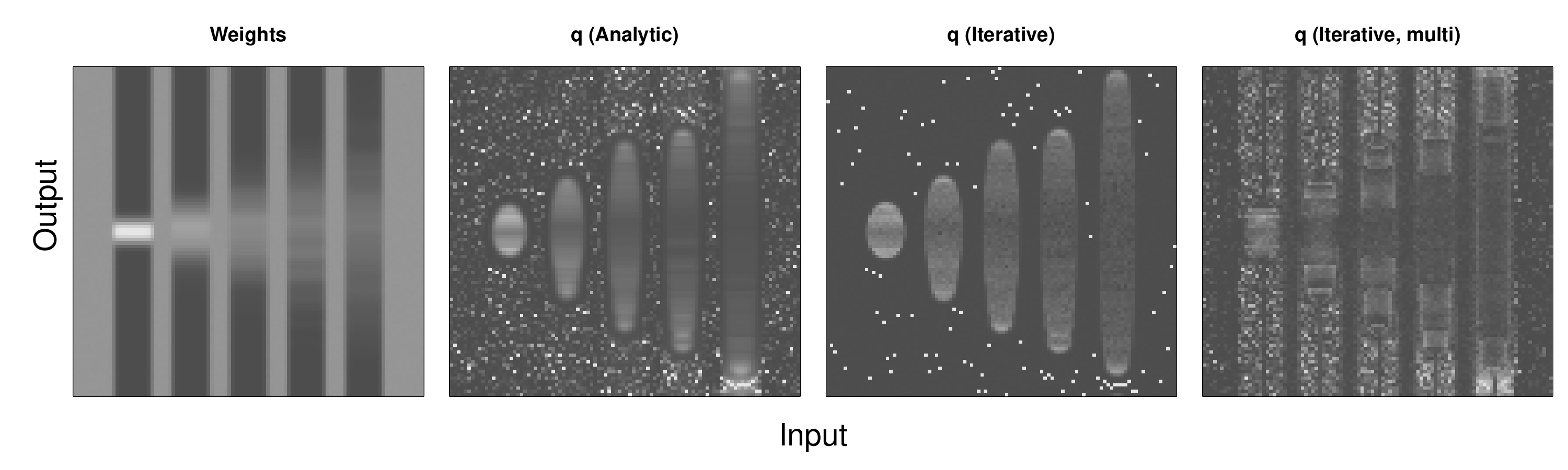}
    \includegraphics[width=\linewidth]{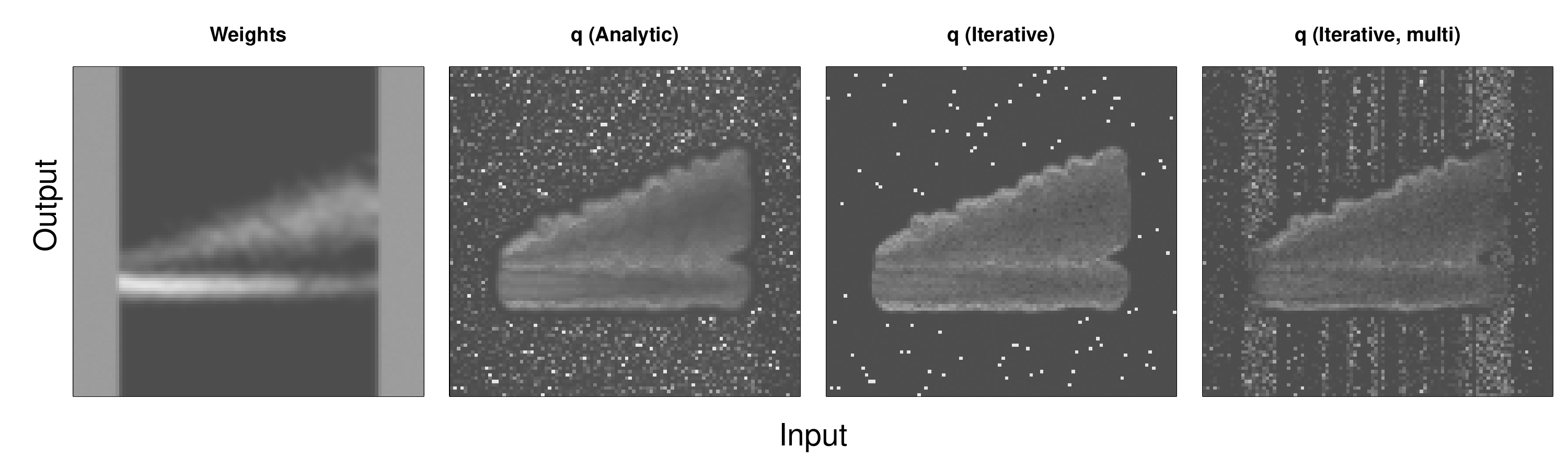}
    \caption{From left to right: (1) Weights relating inputs $\mathbf{X}$ (x-axis) to outputs $\mathbf{Y}$ (y-axis); (2) Analytic release probabilities per input neuron; (3) Release probabilities iteratively learned per neuron; (4) Release probabilities iteratively learned across pooled inputs; Lighter corresponds to higher values in all plots. }
\end{figure}
The matrices of learned weights and their release probabilities from a single iteration of simulation are shown in Figure \ref{fig:weights}, with lighter gray representing higher values in all plots.
In the first panels to the left, the data distributions are clearly encoded by the learned weight values for the active values of the input.
For input domains where no data were simulated, the priors appear as flat, uniform weights over the output domain.
Under all learning schemes for transmission probabilities, the probabilities are highest among weights representing the tails of the distributions.
The heightened transmission probability in the distributional tails serves to counteract the inhibitory effects of the most active output neurons, which would otherwise prevent less likely network states from being sampled.
In coordinates where no data were observed, release probabilities only reflect the random noise of the priors.
The smallest network weights show the same dropout tendencies here as were apparent in the previous simulations of local learning.
Under the analytic mapping, the smallest weights are assigned the highest probabilities. 
Under the iterative, fixed exponent learning rule used here, the smallest weights are neglected in favor of those that most define the observed distribution.
All methods result in the absolute smallest weights transmitting 100\% of the the time, shown as the scattered white pixels, but this aspect is artifactual to our abstracted derivations and is not assigned any practical or theoretical importance.

\section{Discussion}
Our simulations demonstrate that neural networks subject to a range of biological constraints can represent and sample from a complete posterior distribution that includes both observed variation and epistemic uncertainty by synaptic failure alone when release probabilities are derived from weight values. Furthermore, the mapping from weights to release probabilities can be learned locally, using only information available from the present state of the network during each iteration of sampling.

The analytic mapping from weights to release probabilities can be stated concisely: the release probability of each synapse is its strength normalized by the sum over itself and all weaker synapses. 
Intuitively, this means that the normalization factor is largest when there are many projections with just slightly weaker connections.
The strange consequence is that for a nearly uniform conditional distribution, the transmission probabilities will not at all be uniform, but rather hyperbolic.
This distribution is apparent as the sparse noise in the probability matrices shown in Figure \ref{fig:weights} where no data were available.
Further analysis may be warranted to rigorously generalize both our analytic derivation of release probabilities and the local learning rule to the multi-input case, though we found that such rules derived from the single-input simplification appear to work well in simulation.
It is unclear how much tolerance to the inevitable but small approximation errors we should expect from the brain.
For initial analytic steps and notes toward multivariate generalization, see Appendix C. 

\paragraph{Biological plausibility} 
In this paper, we do not comprehensively establish a biological basis of the local learning rule, but only show that such a rule is possible.
\cite{branco2009probability} name several mechanisms that appear to be involved in rapidly modulating dropout rates, including Ca\textsuperscript{2+} ions, astrocytes \citep{semyanov2020making}, postsynaptic endocannabinoids, and other ambient neurotransmitters.
Of particular interest is the similarity of our fixed exponent learning rule, which best balanced simplicity and effectiveness, to the model given by \citep{dodge1967co} in which postsynaptic potential in muscle fibers is related to the fraction of calcium ions bound in the synapse to the fourth power. 
For our learning rule, a power of four is suitable for an accurate distribution over an output layer represented by 10-50 neurons, i.e., projections per input neuron (For more notes and analysis of learning rules, see Appendix B).

We derived a conventional, Bayesian form of epistemic uncertainty for neural networks that maps cumulative synaptic activity to an additional factor in the transmission probabilities.
Our mapping achieves the basic principle of increasing precision as more evidence is gathered.
However, by searching for the simplest mapping or learning rule for epistemic uncertainty, we risk overlooking an important theoretical point.
In the the brain, several advantages may be conferred by situationally modulating epistemic uncertainty.
Whereas observed distributions are, according to our learning rules, shaped relatively slowly over the course of sampling, epistemic uncertainty may be modulated rapidly in response to external contexts and network states more broadly.
In dangerous circumstances, the network may increase transmission probabilities to limit sampling and act swiftly according to only the most likely inferences.
Biological evidence suggests that astrocytes respond to external stimuli and behavior with increased Ca\textsuperscript{2+} signaling by way of neurotransmitters such as noradrenaline, dopamine, and acetylecholine \citep{semyanov2020making, paukert2014norepinephrine}.
Furthermore, hippocampal and cortical astrocytes modulate vesicle release probabilities and plasticity, and may be key to establishing biological control over statistical modes of processing.
In particular, the distal branches of astrocytes undergo rapid, externally induced Ca\textsuperscript{2+} transients as a function of their morphology, which changes in response to local neural activity \citep{semyanov2020making, bazargani2016astrocyte}.
Subsequent learning rules for synaptic failure probabilities should consider mathematical constraints based on the morphological and signaling dynamics of astrocytes at the synapses.
Likewise, our current findings may provide one functional interpretation of such activity.

More broadly, it has been suggested that Ca\textsuperscript{2+} signaling among astrocytes constitutes an additional, complementary pathway for longer-term information processing and modulation of mental states \citep{kastanenka2020roadmap}.
In theory, states of creative problem solving, idle thought, and rumination may be a few examples of posterior-sampling processes that are evoked by restricting Ca\textsuperscript{2+} and broadly reducing vesicle release.
In this mode, as we have shown, the network may conduct searches across its encoded distributions.
By reducing probabilities below the optimal rates for sampling from observed distributions, a network can sample from priors that extend beyond the boundaries of its encoded experiences.
This requires that such priors are true, literal priors, or preexisting synaptic connections that did not result from what we typically regard as learning.

With regard to sampling from observed distributions, the key information used in our local learning rule is the size of the maximum active synapse relative to the sum of all active synapses.
This term is motivated in part by mathematical considerations but is plausible if we assume that the active synapses uptake an extracellular agent at rates related to their sizes. 
The agent may be Ca\textsuperscript{2+} or an ambient neurotransmitter, but most importantly it must only be available in limited supply during each sample.
That way, synapses compete and the amount taken up by each is approximately the aforementioned term in the learning rule: its size or rank relative to the active subset.
Other mechanisms of normalization may exist in the cell bodies of astrocytes with branches extending to the set of relevant, competing synapses in a receiving layer.

\paragraph{Salience and epistemic uncertainty}
In our models, the coefficient $\lambda$ represents the salience assigned to data relative to a regularizing prior.
It is well known that perceptual salience is flexible in the human brain and controlled by many factors, including prior expectations, dangers, and emotional states \citep{kanouse1987negativity, baumeister2001bad}. 
This is unlike most classical statistical analyses in which rows are given equal weight, though inverse probability-weighting and matching schemes are popular among epidemiological research \citep{Rosenbaum87, PaulRubin83, li2018balancing, mansournia2016inverse}.
Under Bayes' theorem, different data points may have different epistemic weight, as each is itself a likelihood with a certain precision corresponding to its perceptual salience. 

\paragraph{Spatial vs temporal representation}
For simple models, population codes may simultaneously represent the complete conditional distribution, making sampling an unnecessarily slow mode of processing.
In more complex models, sampling may be necessary in the same way it is necessary for serial computing: to approximate a complete distribution. 
Population codes provide the capability to fully encode highly complex distributions, but in practice, applying those distributions to a particular action or perception may not be straightforward or possible.
Different intervals of a distribution over one layer may correspond to mutually exclusive endpoints in downstream layers.
Sampling allows the network to control its search across arbitrarily complex domains by engaging in random activity in the simpler, earlier stages of processing leading up to them.
This idea is analogous to sampling from the lowest-dimensional encoding layer of a variational autoencoder \citep{kingma2013auto} to search the final layer of complex reconstructions. 
As the network samples, it can use the relative frequency of downstream activations to weigh the likelihoods or value attributions  of outcomes before making a final selection.

Additionally, resources in the brain may be too limited for complex simultaneous representations.
\cite{levy2002energy} shows that synapses fail at rates that are optimal to promote energy efficiency.
Indeed, it may be natural to link regularization of computational models to resource efficiency of the model architecture, as the motivation behind the former is to produce models with fewer connections and better generalization to out-of-sample data.

These ideas can be further explored by simulations of neural networks with reinforcement learning and appropriately complex, probabilistic tasks.
Future implementations of our model in spiking neural networks are planned, namely in the Axon framework that is currently under development \citep{axonGithub}.
Preliminary tests of an analogous network structure in the same data model presented here have been successful at recovering conditional distributions from spike rates, but more work is needed to further refine and establish our proposed learning rules according to biological findings and limitations.

\subsection{Conclusion}
In summary, this paper demonstrates that the two primary components of Bayesian inference can be represented by population codes and sampled using synaptic failure as the only source of randomness in a network.
We further demonstrate that in a biologically motivated sampling scheme consisting of synaptic failure and lateral inhibition, correct failure probabilities can be learned using only the current, local state of the network during each iteration of sampling.
Rapid modulation of sampling behavior may allow networks to situationally search and evaluate likelihoods over complex, learned distributions, with possible implications for complex planning, problem solving, and creativity.

\bibliographystyle{apalike}
\bibliography{refs}

\newpage

\section{Appendix A: Induction proof for the primary result}
We show that  $p_i = q_i\prod_{j = 0}^{i - 1}(1 - q_j)$ implies $p_i = q_i (1 - \sum_{j = 0}^{i - 1}p_j)$ via induction. 
The base case is trivial: when $i = 1$ then $p_i = q_i \implies p_i = q_i$.
We assume the implication is true for $i = k$, giving us the identity $$p_k = q_k  \prod_{j}^{k - 1}(1 - q_j) \implies p_k = q_k (1 - \sum_{j}^{k-1}p_j)$$ and show that this implies the truth of the statement for $i = k + 1$.
\begin{align*}
	p_{k + 1} &= q_{k+1} \prod_{j}^{k} (1 - q_j) \\
	&= q_{k + 1}(1 - q_k) \prod_{j}^{k - 1}(1 - q_j) \\
	&= q_{k + 1}\lbrace \prod_{j}^{k - 1}(1 - q_j) - q_k  \prod_{j}^{k - 1}(1 - q_j)\rbrace \\
	& = q_{k + 1} \lbrace \prod_{j}^{k - 1}(1 - q_j) - p_k\rbrace \quad \text{ Def'n of $p_k$}\\
	&= q_{k + 1}\lbrace \frac{p_k}{q_k} - p_k \rbrace \quad \text{ Def'n of $p_k$} \\
	&= q_{k + 1} \lbrace1 - \sum_{j}^{k - 1}p_j - p_k\rbrace \quad \text{Induction step} \\
	&= q_{k + 1}(1 - \sum_{j}^{n}p_j) = p_{k + 1} 
\end{align*}

\section{Appendix B: Analysis of learning rules}
The basic form of the learning rule is to iteratively move $\hat q_{i,t}$ toward some target $q^*_{i}$.
$$\hat q_{i,t} = \hat q_{i,t-1} + \gamma(q^*_{i} -  \hat q_{i,t-1}),$$

We defined the true target as:
$$q^*_i = \frac{w_i}{\sum_{j=i}^n w_j}$$

But for local learning, we can use the biased approximation:
$$\hat q^*_{i,t} = \frac{w_i}{\sum_{j\in\mathbf{s}_t} w_j}$$

\paragraph{Mean bias adjustment}
Let us restrict our analysis to the simplest case where $w_i=w_j$, such that $p_i = 1/n$.
Then 
$$\hat q^*_{i,t} = w_i/\sum_{j\in s}w_j = 1/\sum_{j\in s}1 = 1/|\mathbf{s}_t|.$$ 
Multiplying the above estimate by a bias adjustment, the ratio of subset size to total number of weights less than or equal to $w_i$, or $|\mathbf{s}_t|/(n-i+1)$, we get $\hat q_i = 1/(n-i+1) = w_i/\sum_{j=i}^n w_j = q_i$.
Applying the same bias adjustment to non-uniform weights will correct all according to their global mean.
The result is unbiased across weights but highly biased per weight toward their mean value, underestimating the variance and platykurtosis of the modes within the distribution.

\paragraph{Unbiased mean adjustment}
We would like an alternative to the above bias adjustment that preserves variance among the transmission probabilities and avoids bias toward their mean value.
Rather than multiplying the denominator of $\hat q^*_t$ by some value, we can use an exponent.
Intuitively, taking the exponent of a value in $[0,1]$ shrinks smaller values more than larger ones, resulting in exaggerated differences among them:

$$\hat q^*_t = \Big(\frac{w_i}{\sum_{j\in\mathbf{s}_t} w_j}\Big)^{\psi},$$

For uniform weights, the true and approximate targets are just 
$$q^*=\frac{1}{n-i+1},\quad \hat q^* = \frac{1}{|\mathbf{s}_t|}.$$

To simplify the analysis, let us further approximate the local target above with 
$$\hat{|\mathbf{s}_t|}=(n-i)\hat q_{i,t-1}+1.$$

Now, substitute this approximate target into the original learning rule gives us a nonlinear recurrence equation of only $\hat q_{i,t-1}$:
$$\hat q_{i,t} = \hat q_{i,t-1} + \gamma\left( \left(\frac{1}{(n-i)\hat q_{i,t-1}+1}\right)^{\psi} -  \hat q_{i,t-1}\right),$$

For convergence, the second term must equal zero, so we obtain an implicit equation for $\hat q_i$:

$$\left(\frac{1}{(n-i)\hat q_{i,t-1}+1}\right)^{\psi} -  \hat q_{i,t-1} =0,$$

This equation cannot be put in explicit form, but can be plotted or solved numerically
Next, we would like a value of $\psi$ that results in minimal error of approximation to $(\hat q_i - q_i)^2$.
We can find this for the case of uniform weights, because we analytically know the optimal probabilities.
If we assume convergence, such that $\hat q_{i,t} = \hat q_{i,t-1} = \hat q^*_t = \hat q^*_{t-1}$, then 
we can conversely approximate the equilibrium $\hat q_{i,t-1}$ in the above with the analytic solution.
The target may not converge exactly, so we define the approximation error as a function of $\psi$ from the $\mathcal{L}^2$ norm over weight rank $i$:

\begin{align*}
\epsilon\left(\psi\right)&=\int_{1}^{n}\left(\left(\frac{1}{\left(n-i\right)\frac{1}{n-i+1}+1}\right)^{\psi}-\frac{1}{n-i+1}\right)^{2}di, \\
&=\int_{1}^{n}\left(\left(\frac{n-i+1}{2(n-i)+1}\right)^{\psi}-\frac{1}{n-i+1}\right)^{2}di.
\end{align*}

This integral is solvable in closed form but lengthy and complex.
We can numerically or graphically find a minimum value for any weight population size $n$.
Approximation error in the transmission probabilities of the largest weights will have the largest affect on the resulting distribution.
By defining an upper integration boundary $m<n$, we exclude smaller weights and in turn find a value of $\psi$ that better estimates dropout for the larger weights.

The next step may be to take the above integral's closed form and attempt to minimize it with respect to $\psi$, or a function $\psi(n,i)$. 
However, adding complexity to the local learning rule gives us more to explain in biological terms, and constant values of $\psi$ appear to be adequately accurate across values of $n$.
For uniform weights and $n=100$, we obtain $\psi=4.4$, and for $n=1000$, we obtain $\psi=6.6$.
Using only the top $50\%$ of weights in both situations, we get for $n=100, psi=6.25$ and $n=1000, \psi=9.5$.

\paragraph{Subset-dependent $\psi$}
Given that the optimal value of $\psi$ increases with $n$, we may want to modulate it across subsets within a fixed $n$, for example, by using the subset size:
$$\psi_t = |\mathbf{s}_t|,$$

This works because subset size has upper bound $n-i+1$ and an average of approximately 
$$(n-i)\mathbb{E}[q_{j,t\in \mathbf{s}_t\setminus \{i\} } ]+1,$$ 
so the penalty is smallest on average for small weights as a function of fewer possible subset members with average probability of transmission increasing as $i\to n$.
As the number of weights $w_j < w_i$ is distributed approximately as a Binomial over samples, we can eliminate unnecessary noise from $\psi_t$ by using only the mean value of $|\mathbf{s}_t|$ given $\hat q_{i,t-1}$ and the index of the largest weight, $i$:
$$\psi_t = (n-i)\hat q_{i,t-1}+1,$$

Alternatively, the following set size expectation with an additional adjustment appears to provide even better results:
$$\sqrt{\hat q_{i,t-1}}((n-i)\mathbb{E}[q_{j,t\in \mathbf{s}_t\setminus \{i\} } ]+1),$$ 

In any case, consider one possible learning rule:
$$\hat q_{i,t} = \hat q_{i,t-1} + \gamma\Big[ \Big( \frac{w_i}{\sum_{j\in\mathbf{s}_t} w_j}\Big)^{(n-i)\hat q_{i,t-1}+1} -  \hat q_{i,t-1}\Big].$$

This learning rule is a coupled dynamic system in which the parameter, $\hat q_{i,t}$ and its target $\hat q^*_t$ share a mutual attractor that is close to the true target, $q^*$.
The stochastic component of this system is the denominator of $\hat q^*_t$, summing over the subset of weights.
The ratio of each weight to the others present in the subset is the critical information for an accurate solution, and cannot be approximated to further reduce stochasticity.
Neither term converges to the attractor but the final estimate can be recovered either by $\mathbb{E}[\hat q_{i,t}]$ or $\mathbb{E}[\hat q^*_t]$, or by decaying $\gamma \to 0$ at an appropriate rate and number of iterations. 

\section{Appendix C: Dropout under multiple inputs}
In our simplified derivation of dropout probability per weight, we constrain the problem to $y_i=w_i$. But more generally, $\mathbf{y}=\mathbf{W}^\intercal \mathbf{x}$, and for large $n$, we can approximate the distribution $y_i \sim \text{Binomial}(|\mathbf{x}|, \mathbb{E}\left[\mathbf{W}_i^\intercal \mathbf{x}\right])$.
We must find dropout probabilities such that,
$$\mathbf{W}_i^\intercal \mathbf{x} = P(y_i > y_j),\quad \forall j \neq i$$
A useful precedent is given by \cite{mostafa2018learning}, in which they approximate the SoftMax function with the probability that each activation in the receiving layer, Gaussian distributed at the limit of input layer size, is higher than the activation with the largest mean value.
We would therefore need to solve for dropout probabilities such that the softmax approximation for each receiving neuron $i$ is as close as possible to normalized $\mathbf{W}_i^\intercal \mathbf{x}$.
That approximation is,
$$\hat p_i = \underset{j\neq i}{\min}\left\{ \text{cdf}\left( \frac{\mu(y_i)-\mu(y_j)}{\sqrt{\sigma^2(y_i)+\sigma^2(y_j)}}; 0, 1 \right) \right\} 
= \frac{\mathbf{W_i^\intercal x}}{\sum_i^n \mathbf{W_i^\intercal x}},$$
the right-hand side of which translates to the minimum probability that output $i$ is greater than any other output $j$.
Alternatively, we can say that the bracketed term is the unnormalized probability that neuron $i$ out-competes the neuron with the largest average activation.
If we substitute in moments from the scaled Binomial distributions and fix $j$ to represent the neuron with the largest activations on average, we get
$$\hat p_i = \text{cdf}\left( \frac{\mathbf{ (W_i \odot  Q_i)^\intercal \mathbf{x}}-(\mathbf{W_j\odot Q_j)^\intercal \mathbf{x}}}{\sqrt{\mathbf{(W_i^2 \odot  Q_i \odot (1-Q_i))^\intercal x^2 }+\mathbf{ (W_j^2\odot Q_j\odot (1-Q_j))^\intercal x^2}}}; 0, 1 \right)
= \frac{\mathbf{W_i^\intercal x}}{\sum_i^n \mathbf{W_i^\intercal x}}.$$

The probability of each neuron winning is given by the value of a cumulative normal distribution at the  normalized difference of that neuron from the max active neuron.


\end{document}